%% file: manuscript.tex
\documentclass[lettersize,journal]{IEEEtran}
\pdfoutput=1
\usepackage{algorithm}
\usepackage{algorithmic}
\usepackage{bbding,pifont}
\usepackage{amsthm,amsmath,amssymb}
\usepackage{graphicx}
\usepackage{subfig}
\usepackage{siunitx}
\usepackage[colorlinks,linkcolor=red]{hyperref}
\usepackage{color}
\usepackage[noadjust]{cite}
\usepackage{xcolor}
\usepackage{ulem}
\usepackage{pifont}
\usepackage{multirow}

\begin{document}

\title{Advancing Radar Hand Gesture Recognition: A Hybrid Spectrum Synthetic Framework Merging Simulation with Neural Networks}

\author{ Jiaqi~Tang,~Xinbo~Xu, Yinsong~Xu,~and~Qingchao~Chen\ding{41}
\thanks{This work was supported by the grants from the National Natural Science Foundation of China (62201014), Clinical Medicine Plus X - Young Scholars Project of Peking University, and the Fundamental Research Funds for the Central Universities.}
\thanks{Jiaqi Tang and Qingchao Chen are with the National Institute of Health Data Science, Peking University, Beijing, 100191, China. They are also affiliated with the Institute of Medical Technology, Peking University, and the State Key Laboratory of General Artificial Intelligence, Peking University, Beijing, China. (e-mail: jiaqi.tang818@gmail.com, qingchao.chen@pku.edu.cn)}
\thanks{Xinbo Xu is with the School of Computer Science, Beijing University of Posts and Telecommunications, Beijing 100876, China. (e-mail: ZnPt@bupt.edu.cn)}
\thanks{Yinsong Xu, Beijing University of Posts and Telecommunications, Beijing 100876, China. He is also affiliated with the National Institute of Health Data Science, Peking University. (e-mail: xuyinsong@bupt.edu.cn)}
\thanks{\ding{41} \textit{Corresponding author: Qingchao Chen(email:qingchao.chen@pku.edu.cn)}}
}

\maketitle

\input{abstract}

%

\input{introduction}

\input{related_work}


\input{methods}

\input{experimental_setup}

\input{results}

\input{conclusion}




\ifCLASSOPTIONcaptionsoff
  \newpage
\fi



%
\bibliographystyle{IEEEtran}
\bibliography{cite}



%








\end{document}

%% file: abstract.tex
\begin{abstract}
Millimeter wave (mmWave) radar sensors play a vital role in hand gesture recognition (HGR) by detecting subtle motions while preserving user privacy. However, the limited scale of radar datasets hinders the performance.  Existing synthetic data generation methods fall short in two key areas. On the one hand, modeling-based approaches fail to accurately simulate the wave propagation and reflection at the hand-gesture level, facing unique complexities such as diffraction and occlusion. On the other hand, generative model-based methods are hard to converge while radar data is limited, lacking interpretability, and sometimes fail to produce kinematically plausible results.
To overcome these limitations, we propose a novel hybrid spectrum synthetic framework leveraging visual hand gesture data. It combines a cylinder mesh-based hand reflection model with a small-scale neural network called RadarWeightNet, which focuses on assigning weights to simulated signals. Our framework addresses two key challenges: achieving accurate simulation of complex hand geometry and bridging the simulation-to-real gap in a data-driven manner while preserving interpretability, which balances physical accuracy with machine learning adaptability.
We tested our framework under extreme scenarios where radar data is scarce. The results demonstrate the effectiveness of our hybrid framework, achieving up to 63\% SSIM in synthetic performance and up to 30\% improvement in classification performance in few-shot learning.
\end{abstract}


%% file: introduction.tex
\section{Introduction}


\IEEEPARstart{W}{ith} the development of the Internet of Things and ubiquitous sensing technologies, Human-Computer Interactions (HCI) have gained unprecedented attention and significance. Among the various HCI modalities, Hand Gesture Recognition (HGR) has emerged as a pivotal interface mechanism, with substantial applications in Virtual Reality (VR)~\cite{xu2006neural}, Augmented Reality (AR)~\cite{liang2015ar}, and contactless vehicle control systems~\cite{francis2014significance}. To date, HGR is facilitated primarily by four types of sensors: cameras~\cite{suarez2012hand}, acoustic devices~\cite{wang2020push}, wearable sensors~\cite{al2022multi}, and radar systems.
Camera-based sensors for human gesture recognition (HGR) risk privacy breaches, acoustic sensors require close proximity to the hand and are more easily affected by ambient noise, which may not be practical for all applications, and wearables are prone to user discomfort and environmental interferences. In contrast, Frequency-Modulated Continuous Wave (FMCW) radar offers a compelling alternative. It is cost-effective and proficient in discerning subtle hand motions through the analysis of time-delay and Doppler-shifted signals~\cite{gurbuz2019radar}.

Current state-of-the-art approaches for HGR leverage deep learning algorithms for the classification of radar spectrum~\cite{skaria2019hand,liu2019spectrum,lien2016soli,skaria2020deep}. Despite these advances, the performance of radar-based HGR systems lags, particularly when contrasted with video-based methods~\cite{molchanov2016online,gadekallu2021hand,kopuklu2019real}. The challenges faced by radar-based HGR systems can be delineated as threefold: firstly, the limited scale of radar datasets, which are costly and laborious to collect and annotate~\cite{tang_fmnet_2022}; secondly, radar data augmentation proves difficult because each Doppler bin's velocity information is critical, rendering traditional image augmentation techniques like cropping and rotation ineffective; thirdly, the variability introduced by factors such as environmental conditions, sensor specifications, and target attributes complicates the training of deep learning models for radar signals~\cite{vishwakarma_simhumalator_2022}.

Existing synthetic spectrum generation methodologies, broadly classified into modeling-based and generative model-based approaches, have significant limitations when applied to hand gesture recognition.  \textbf{ On the one hand, modeling-based methods \cite{ram2008simulation, vishwakarma_simhumalator_2022} mainly focus on the human body level and fail to capture the intricacies of hand structures.} The similarity between finger size and millimeter-wave radar wavelength leads to complex diffraction and multipath interference issues, which are more problematic than full-body modeling. Moreover, the high degree of freedom in hand movements, coupled with frequent self-occlusion and the radar's limited spatial resolution, further complicates accurate simulation. 
\textbf{On the other hand, generative model-based approaches\cite{ doherty2019unsupervised, rahman_physics-aware_2021, ahuja_vid2doppler_2021}, are heavily dependent on large quantities of real radar data, struggling to converge when applied to the limited datasets(refer to Section V.A.), which is common in HGR applications. Also, it lacks interpretability and produces data that lack kinematic constraints. }
These limitations highlight the need for a more sophisticated approach that can accurately model the complexities of hand gestures while overcoming the constraints of limited data availability.


To address these challenges, we propose an innovative hybrid spectrum synthetic framework, integrating
the strengths of two methods, guided by the following two key design principles: (1) Accurately modeling the complex geometry and kinematics of the human hand, (2) Bridging the simulation-to-reality gap through data-driven optimization on signal level while preserving good interpretation and following kinematic constraints. Our approach combines physically-based modeling with machine learning adaptability to overcome the limitations of existing methods. 
The framework consists of two main components:
\textbf{
(1) A fine-grained cylinder mesh-based hand model that accurately captures the complex anatomy and motion of the human hand, which leverages visual data from various hand gesture vision modalities. (2) RadarWeightNet, a neural network designed to optimize the final simulated radar signals by weighing each reflection point's contribution.}

Our framework begins with the simulation of intermediate frequency(IF) signals of each reflection point based on the hand model using FMCW radar equations, accounting for critical parameters such as propagation time delay, radar cross section(RCS), and transmission power from the hand model and the radar settings. These simulated IF signals are then refined by RadarWeightNet, which takes as input various motion characteristics derived from 3D coordinates, including the number of visible vertices, velocity, and acceleration of hand segments. Unlike the traditional modeling-based methods that sum all the IF signals together to generate the final one, our RadarWeightNet predicts weights for each reflection point's IF signal thus effectively fine-tuning the synthetic signal to account for complex phenomena such as occlusions, multipath effects, and subtle interactions between reflection points. The final synthetic time-doppler spectrum is produced by applying a Short-Time Fourier Transform (STFT) to the weighted signal. To ensure high fidelity, the entire process is optimized using the Structural Similarity Index Measure (SSIM) as a loss function, which ensures that the synthetic spectrums closely match real ones. This integrated approach combines physically-based modeling with data-driven optimization, enabling the generation of highly realistic radar spectra for hand gestures and effectively addressing the challenges posed by limited radar datasets in hand gesture recognition.

To rigorously evaluate our framework, we developed a comprehensive multi-modal dataset comprising 10 distinct hand gestures captured by calibrated mmWave radar and an infrared camera. This dataset encompasses varying gesture complexities, occlusion scenarios, and angle variabilities, including both single and dual-handed actions across multiple incidence angles (0°, 30°, and -30°), detailed in Section IV. This extensive dataset enabled us to demonstrate our framework's capability to generate synthetic spectrum closely resembling real radar data, with RadarWeightNet further significantly enhancing this similarity. We then tested our framework's robustness and its ability to improve HGR accuracy, achieving up to a 30\% improvement in classification performance under few-shot learning conditions. Furthermore, we validated our approach by introducing an external vision dataset and pre-training the classification model using the corresponding synthesized spectrums, thereby demonstrating the framework's effectiveness in leveraging external data sources to enhance radar-based gesture recognition.

To sum up, the main contribution of this article can be summarized as follows.
\begin{enumerate}
    \item We captured a comprehensive radar-vision HGR dataset, which is uniquely benchmarked across three dimensions: occlusion, angle, and number of hands enabling a thorough evaluation of synthetic radar data.

    \item We propose a hybrid synthetic framework that integrates the radar hand reflection model, and RadarWeightNet, bridging the simulation-to-real gap in a data-driven manner while preserving physical interpretability.

    \item The synthetic framework can leverage external vision datasets to augment radar datasets and significantly improve classification results in a few-shot manner.
\end{enumerate}

%% file: related_work.tex
\section{Related Work}



\begin{figure*}[t]
    \centering
    \includegraphics[width=0.99\linewidth]{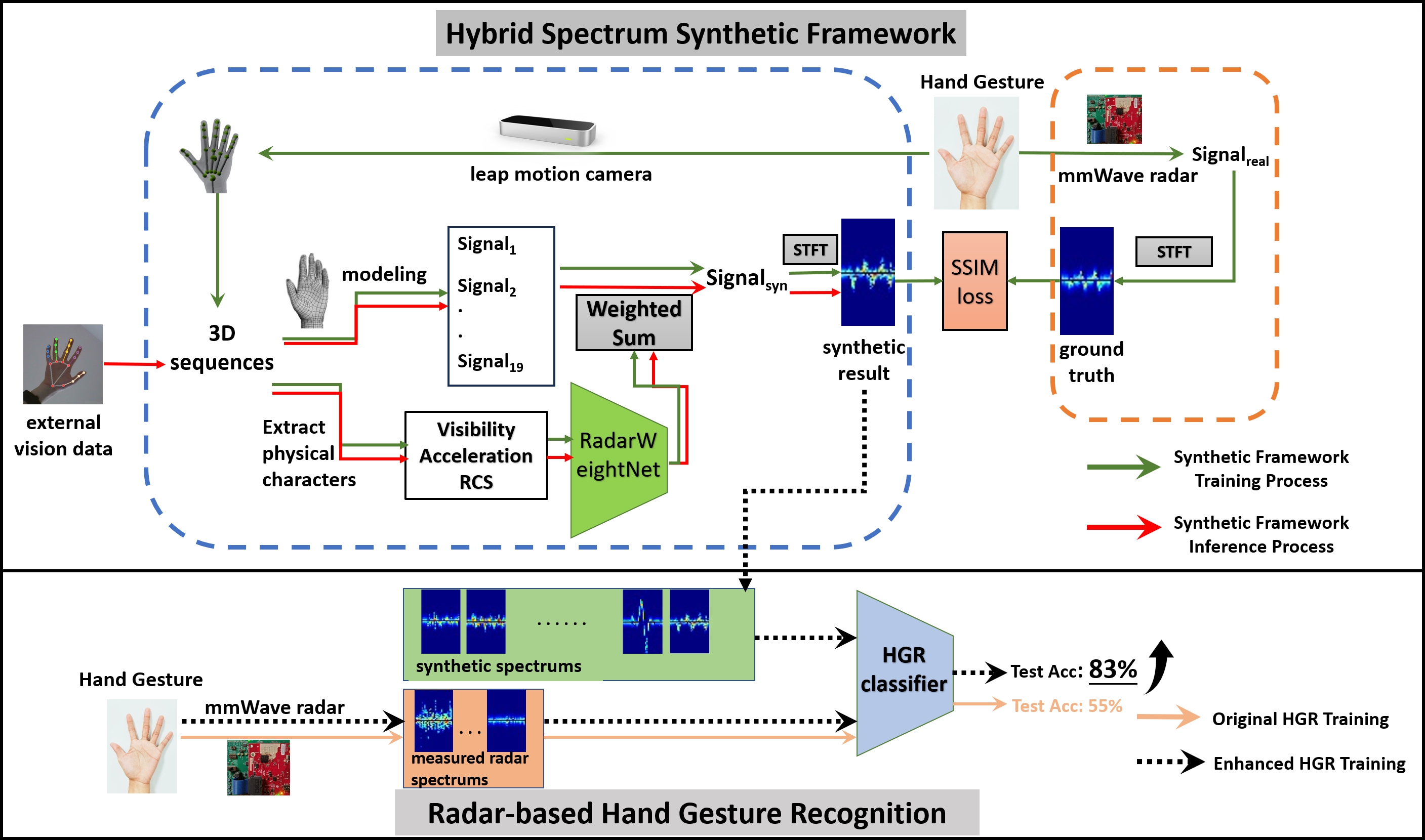}
    \caption{The system overview of our hybrid synthetic framework and how it enhances the HGR accuracy by easing the dataset scale limitation. The training process of the synthetic framework (green path) uses paired mmWave radar data and 3D hand coordinates to optimize RadarWeightNet, which reweights modeled radar signals from a cylinder mesh-based hand model. The inference phase (red path) generates synthetic radar spectrums from external vision modality and then expands the spectrum dataset, improving the classification performances in radar-based hand gesture recognition.}
    \label{fig:system architecture}
\end{figure*}

\subsection{Hand Gesture Recognition Based on FMCW Radar}
Recent efforts in radar-based hand gesture recognition (HGR) have explored extracting micro-movements from raw radar signals through various data representations \cite{lien2016soli, li2017sparsity, hazra2018robust}.
A pioneering project is Soli \cite{lien2016soli}, which utilized an FMCW radar and applied a random forest algorithm to accurately classify four hand gestures. Building on this, Li \textit{et al.} \cite{li2017sparsity} introduced a sparsity-driven approach that processes time-doppler features from radar echoes for input into diverse classifiers. Subsequently, Hazra and Santra \cite{hazra2018robust} presented a 60-GHz mm-wave radar sensor for short-range applications, employing a Long Recurrent CNN architecture to extract features from range-doppler heatmaps.
Ahmed \textit{et al.} \cite{iotjmultistream} integrated multiple data representations into a multistream CNN architecture, leveraging concatenated convolutional features from range-time maps, Doppler-time maps, and angle-time maps to enhance classification performance.
Further advancing the field, Zhang \textit{et al.} \cite{zhang2021xgest} proposed a knowledge transfer framework to classify new, unseen gesture classes by utilizing diverse gesture datasets. Addressing the challenge of limited datasets and potential overfitting, a meta-learning framework was introduced by Shen \textit{et al.} \cite{shen2022metalearning} that capitalizes on prior domain knowledge for swift adaptation to novel hand gesture classes.

While these studies have made significant advancements by innovating neural network architectures and loss functions for enhanced feature representation, the critical issue of dataset scarcity remains largely unaddressed. \textbf{Unlike the incremental improvements on neural network architectures and loss functions, our method aims to fundamentally enhance the training data through a hybrid mmWave radar simulation framework coupled with enhanced neural network fine-tuning. This not only provides a richer dataset for training more robust models but also sets a new direction for future research in the radar HGR domain.}

\subsection{Radar Spectrum Synthesis}
Synthetic approaches for FMCW radar spectrum can be categorized into two primary groups: modeling-based methods and deep learning-based generative models.

Modeling-based synthetic methods are based on radar system parameters alongside the physical and motion characteristics of the target, conforming to the radar equation \cite{ram2008simulation, chipengo_high_2021, vishwakarma_simhumalator_2022, ninos2021synthetic}. Ram and Ling \cite{ram2008simulation} presented a pioneering Doppler radar synthetic data generator that produces micro-Doppler signatures from various human movements using computer animation data, calculating the radar cross-section (RCS) at each time step to simulate human activity patterns. Vishwakarma \textit{et al.} \cite{vishwakarma_simhumalator_2022} advanced this concept with a data-driven approach, harnessing a motion capture suit to extract precise positional data from video frames, thus allowing for the adjustment of radar, target, and signal processing parameters, which is a step beyond the method presented in \cite{ram2008simulation}. While these modeling-based synthetic methods have shown promise for full-body human motion simulation, they face significant challenges when applied to hand gesture recognition. The complexity of hand structures, with their small, intricate components comparable to millimeter-wave radar wavelengths, introduces difficulties in accurately modeling diffraction and occlusion effects.

In contrast, deep learning methods have been leveraged to construct generative models for synthesizing radar signals \cite{ishak2018human, erol_gan-based_2019, chipengo_high_2021}. Rahman \textit{et al.} \cite{rahman_physics-aware_2021} proposed a multi-branch GAN architecture that incorporates a physics-aware perspective into the network, enhancing the kinematic realism of the generated data. Ahuja \textit{et al.} \cite{ahuja_vid2doppler_2021} introduced a novel encoder-decoder architecture that translates video recordings of human activities into synthetic Doppler radar data. Additionally, Zhao \textit{et al.} \cite{zhao_synthetic_2022} and Tang \textit{et al.} \cite{tang_fmnet_2022} have developed methodologies for synthetically replicating radar signal features to aid in the training of classification networks. However, these methods require large quantities of radar data in the beginning, which makes the model struggle to converge when data is scarce. Moreover, the lack of physical interpretability and reliability potentially violates hand movement constraints and fails to capture complex radar-hand interactions.

Our work distinguishes itself by addressing the limitations of both modeling-based and generative model-based approaches through a novel hybrid mmWave radar simulation framework. This framework combines a fine-grained cylinder-based hand model for accurate simulation of complex hand geometry and occlusion with RadarWeightNet, an MLP-based neural network that fine-tunes simulated signals. By integrating physically grounded simulation with data-driven optimization, our approach overcomes the challenges of accurately modeling hand-level complexities while preserving physical interpretability and kinematic constraints.

%% file: methods.tex
\section{Methods}

\subsection{Overview}


Current radar spectrum synthesis methods for hand gesture recognition face significant limitations. On the one hand, modeling-based approaches struggle to accurately simulate the complex wave propagation and reflection at the hand-gesture level, particularly due to challenges like diffraction and occlusion arising from the similarity between finger size and millimeter-wave radar wavelength. On the other hand, generative model-based methods often fail to converge with limited radar data, lack interpretability, and may produce kinematically implausible results.
To address these challenges, we propose a novel hybrid spectrum synthetic framework to leverage the abundant visual hand gesture dataset by combining two key components: (1) a fine-grained cylinder mesh-based hand reflection model to simulate initial radar signals for each part of the hand. and (2) RadarWeightNet, a small-scale neural network to optimize the simulated radar signals by predicting weights for each reflection point.

As illustrated in Fig \ref{fig:system architecture}, our hybrid spectrum synthetic framework operates in two distinct phases: training and inference. During the training phase, we simultaneously collect ground truth radar spectrums using mmWave radar and 3D hand joint coordinates via a Leap Motion controller. These 3D coordinates serve a dual purpose: they are input into our cylinder-based hand model to generate initial radar signals, and they are processed by RadarWeightNet to extract motion features and compute weights for different reflection points. The initial signals and computed weights are then combined to produce a synthetic final signal, which is used to generate a spectrum. This synthetic spectrum is subsequently compared against the ground truth to optimize RadarWeightNet. In the inference phase, we leverage the trained system to generate synthetic radar spectrums from any vision modality. This process involves aligning hand skeleton sequences from the vision dataset with our hand model, generating initial radar signals based on these aligned skeletons, and applying the trained RadarWeightNet to compute appropriate weights for each reflection point. The final synthetic radar spectrum is produced by combining these weighted signals. 

The following subsections detail the key components of the proposed hybrid spectrum synthetic framework. Section III.B. outlines radar signal preprocessing, explaining how to obtain the measured radar spectrum. Section III.C. introduces the cylinder mesh-based model for calculating wave propagation and reflection, used for signal simulation. Section III.D. explains RadarWeightNet, focusing on its design principles and network structure, and also explains how it cooperates the hand reflection model and balances physical accuracy with machine learning adaptability, enabling the generation of realistic radar spectrums for hand gestures while preserving interpretability and kinematic constraints.




\subsection{Radar Signal Preprocessing}


\begin{figure}[h]
    \centering
    \includegraphics[width=3.5in]{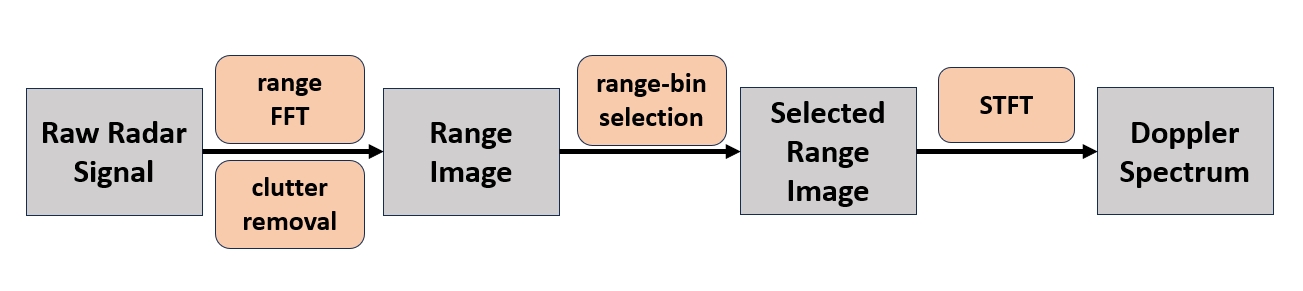}
    \caption{Radar signal processing pipeline.}
    \label{fig:signal_processing}
\end{figure}

Frequency-Modulated Continuous Wave radar works by emitting continuous waves. After the modulated signal is transmitted toward a target object and reflected back, the received signal is then mixed with the transmitted signal to obtain an intermediate frequency (IF) signal. We could get the range and velocity information from the frequency and phase change of the signal. 

As shown in Fig. \ref{fig:collection_system}, we first apply the Range-FFT to get the distance of the reflecting object and select the appropriate range bin where the hand gesture was done. The distance information is contained in the time delay of the reflected signal with respect to the transmitted signal. The relationship between the frequency of IF
signal f and the distance d between radar and object can
be denoted as \cite{li2022towards}
\begin{equation}
f = S \cdot \tau =\frac{S\cdot2d}{c} \to  d= \frac{fc}{2S}
\label{eq:distance_calc}
\end{equation}
where $S$ is the slope of the chirp signal, $\tau$ is the time delay between the transmitted and received signals, and $c = 3 \times 10^8 m/s$ is the speed of the signal. To remove the clutter in the background, static clutter is suppressed by subtracting the mean values after the range-FFT.

Then to extract the velocity information, we need to analyze the time-varying frequency shift in the selected range bin, which could be calculated from the phase difference between the transmitted and received signals. Specifically, the phase difference can be expressed as:
\begin{equation}
\Delta\phi=2\pi\Delta f\tau=\frac{4\pi v T_c}{\lambda} \to v=\frac{\lambda\Delta\phi}{4\pi T_c}
\label{eq:velosity_calc}
\end{equation}
where $v$is the velocity of the target , $\Delta\phi$ is the phase difference, $\Delta f$ is the frequency shift, and $\lambda$ is the wavelength of the signal.

Based on the equation above, we compute the Short-Time Fourier Transform(STFT) spectrograms by utilizing the method from \cite{chen2019micro}.  Specifically, STFT is applied for each channel using the Hanming windows. The FFT window length is set to 64 points, and the step is set to 64 time samples to get the spectrum with the size of 64 × 32 pixels, which are then
used for classification.

\subsection{Cylinder Mesh-based Hand Reflection Model}

This subsection details the Cylinder Mesh-based Hand Reflection Model, a crucial component of our synthetic framework. First, we explain the hand motion acquisition process using the Leap Motion controller to capture precise hand movements. Next, we describe the creation of a 3D Hand Mesh Based on a Cylinder-shaped Skeleton, transforming the captured data into a detailed mesh model. Finally, we outline the signal synthesis from reflecting points, demonstrating how radar signals are generated based on this reflection.

\begin{figure*}[t]
    \centering
    \includegraphics[width=1\linewidth]{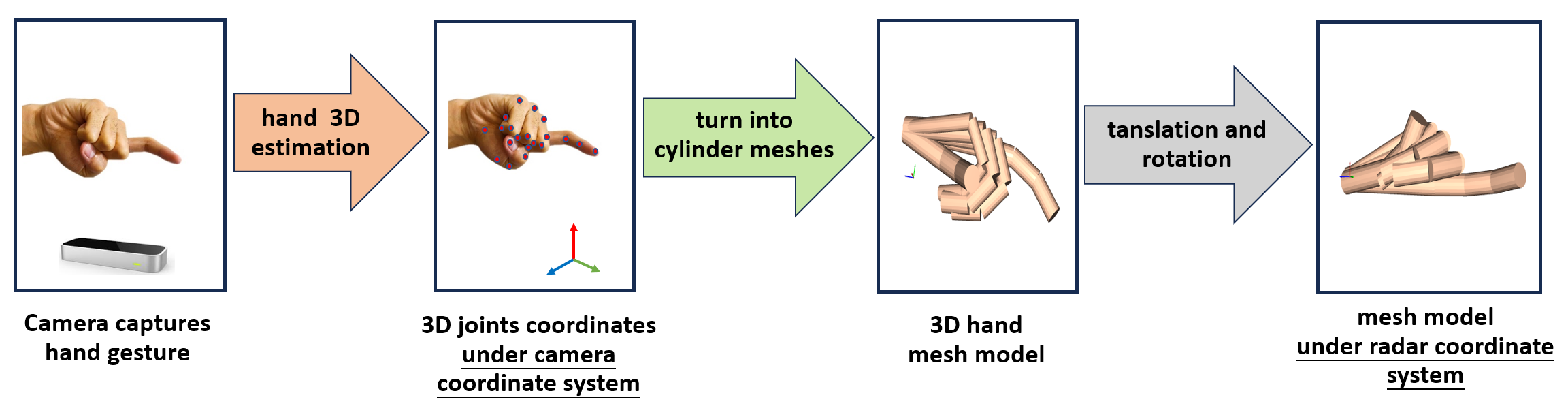}
    \caption{Hand modeling pipeline overview. A camera captures hand gestures and estimates 3D joint coordinates, which are used to construct a cylinder-based 3D mesh model of the hand. This model accurately represents hand geometry and occlusion, and is transformed into the radar coordinate system. The resulting detailed hand model enables the simulation of radar reflections from multiple scattering points, allowing for realistic radar signal synthesis that accounts for complex hand structures and movements.} 
    \label{fig:simulation_model}
\end{figure*}

\subsubsection{\textbf{Hand Motion Acquisition}}

The Leap Motion controller \cite{bachmann2018review} offers a tool for capturing hand movements and postures with high precision, transforming them into skeletal data. Specifically, the thumb consists of three parts, while the remaining four fingers each consist of four parts: the metacarpals (excluding the thumb), proximal phalanges, intermediate phalanges, and distal phalanges. At any given instant, the Leap Motion Controller is capable of capturing the 3D coordinates of both ends of these phalanges, ultimately creating a detailed representation of 20 joints within the Leap Motion controller's coordinate system.
Since the radar antenna and Leap Motion controller are placed in different locations, we need to transform coordinates to the radar-centered system, thereby enabling precise tracking and further analysis. First, we ensure that both the radar antenna and the Leap Motion controller are positioned on the same horizontal plane, with their sensing directions pointing upwards. We then precisely measure the horizontal distance between the center of the radar antenna and the center of the Leap Motion controller's lens, obtaining the offset in the X and Y directions.
Based on these measurements, we apply a simple translation to the Leap Motion coordinates:

\begin{equation}
\begin{aligned}
X_{radar} &= X_{leap} - \Delta x, \\
Y_{radar} &= Y_{leap} - \Delta y, \\
Z_{radar} &= Z_{leap},
\end{aligned}
\end{equation}

where $(X_{leap}, Y_{leap}, Z_{leap})$ are the coordinates from the Leap Motion controller, $(X_{radar}, Y_{radar}, Z_{radar})$ are the corresponding coordinates in the radar-centered system, and $(\Delta x, \Delta y)$ is the measured offset between the two devices. This transformation enables precise tracking and further analysis in the radar-centered coordinate system.

\subsubsection{\textbf{3D Hand Mesh Based on Cylinder-shaped Skeleton}}


Distinct from the common practice of using ellipsoids to model various body parts, we have adopted a cylinder-based approach to effectively capture the unique morphological characteristics of the human hand. Our model represents the phalanges of the hand using cylinders of varying sizes, assuming that the radar scattering centers are located at the center of the bones. Consequently, the model includes 19 scatterers to represent the human hand.
The methodology of this simulation is visually depicted in Fig. \ref{fig:simulation_model}. To transform the skeletal data into a cylinder-based mesh, we follow the following process:
Firstly, we determine the position of a point on the wrist, which is used as the origin. Then, the positions of all other skeletal points are generated relative to the direction vector of the starting point, with distances corresponding to the lengths of the standard hand's bones.
Upon generating this skeleton model, the bones are replaced by cylinders, with the top and bottom surfaces of these cylinders aligned with the ends of the respective bones. In our framework, each cylinder is represented by 421 vertices. Specifically, a circle is represented by 20 vertices, the sides are represented by 20 circles, and an additional two vertices are located at the center of each of the upper and lower bottom surfaces.

\subsubsection{\textbf{Signal Synthesization from Reflecting Points}}

According to \cite{zhang_synthesized_2022}, the transmitted signal $S_{TX}$ of FMCW at the time t could be expressed as:
\begin{equation}
    S_{TX}(t) = Ae^{j(2\pi (f_0t+\frac{B(t^2_m+mT^2)}{2T}))+\phi_0}
    \label{eq:tx_signal}
\end{equation}
where
$A$ is the amplitude of the transmitted signal.
$j$ is the imaginary unit.
$f_0$ is the carrier frequency of the transmitted signal.
$B$ is the bandwidth of the transmitted signal.
$t_m$ represents the time delay of the received signal relative to the transmitted signal.
$T$ is the pulse duration or the time between transmitted pulses.
$m$ is an integer that represents the number of transmitted pulses.
$\phi_0$ is the initial phase of the transmitted signal.
When the transmitting signal bounces off a reflection point at distance $D$ with respect to the receiving antenna, ignoring the multipath and other electronic interference, the received signal is simply the attenuated, time-delayed, and Doppler-shifted, versions of the transmitted signal\cite{vishwakarma_simhumalator_2022} which could be expressed as:
\begin{equation}
    S_{RX}(t) = A'e^{j(2\pi (f_0(t-\tau)+\frac{B((t_m-\tau)^2+mT^2)}{2T}))+\phi_0}
    \label{eq:rx_signal}
\end{equation}
where $A'$ is the attenuated amplitude, which could be obtained according to the radar communication principle \cite{instruments2017introduction}:
\begin{equation}
    A' = \frac{\lambda \sqrt{ G_{Tx}G_{Rx} P\sigma }}{(4\pi)^{1.5}D^2}
    \label{eq:attenuated_amplitude}
\end{equation}
where 
$\lambda$ is the wavelength of the transmitted signal, $G_{Tx}/G_{Rx}$ represents the antenna gain of the transmitting and receiving antenna respectively. $\sigma$ is the radar cross section(RCS), and $P$ is the transmission power.
For the estimation of non-normal incidence backscattered RCS of a circular cylinder in Equation \ref{eq:attenuated_amplitude} has been investigated and included in the RCS handbook \cite{RCS} before as follows:
\begin{equation}
    \sigma = \frac{\lambda r \sin{\theta}}{8 \pi (\cos{\theta})^2}
    \label{eq:cylinder_rcs}
\end{equation}
where  $r$ is the radius of circular cylinder, and $\theta$ is the aspect angle from radar's viewpoint.

However, this simplified model does not fully capture the complexities of real-world scenarios, where factors such as occlusions, multi-path effects, and scattering and diffractions can significantly influence the radar signals. To address these challenges and enhance the accuracy of our synthetic data, we introduce a novel deep-learning architecture called RadarWeightNet.

\subsection{RadarWeightNet Details}

To address the limitations of the simplified model in capturing real-world complexities, we introduce RadarWeightNet. we maintain the integrity of the IF signals simulated by our cylinder mesh-based hand reflection model. RadarWeightNet solely reweights these signals without any further transformation, ensuring that critical properties such as frequency shifts and time delays are preserved. This approach maintains a strong physical grounding by directly linking each reflection point on the hand model to its contribution to the final radar signal.  The network outputs a set of weights that modulate the importance of individual reflection point signals, enabling fine-grained adjustment of synthetic signals. This approach effectively compensates for potential modeling errors and captures subtle interactions between reflection points that may be overlooked in purely physical models. RadarWeightNet learns to adjust each reflection point's contribution based on its spatial relationships and motion characteristics within the hand gesture, achieving a balance between physical accuracy and data-driven adaptability.

\begin{figure}[htbp]
    \centering
    \includegraphics[width=3in]{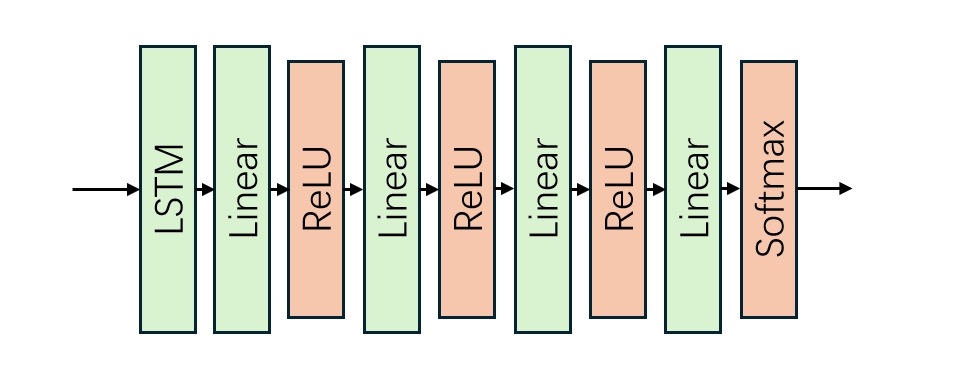}
    \caption{Detailed architecture of RadarWeightNet.}
    \label{fig:network_architecture}
\end{figure}

The architecture of RadarWeightNet, as shown in Fig .\ref{fig:network_architecture}, comprises an LSTM layer to capture temporal dependencies and three fully connected layers for non-linear transformations and dimensionality reduction.
RadarWeightNet takes as input a set of motion characteristics observed from the radar perspective, including the number of visible vertices, velocity, and acceleration. This unique input strategy allows the network to implicitly extract information about the relationships among different reflection points, considering their dynamic characteristics during motion. The processing pipeline involves feeding these motion characteristics into the network, which then predicts weights for each reflection point. We calculate the final signal by applying these weights to the original reflection point signals, then apply the Short-Time Fourier Transform (STFT) to generate the synthetic spectrum. Importantly, throughout this process, the original signals from each reflection point remain unaltered, adhering to basic kinematics and helping avoid systemic errors that might arise in purely generative models. The synthetic spectrum is compared with the real spectrum using the Structural Similarity Index Measure (SSIM) as the loss function for optimization. loss function, which is defined as:

\begin{equation}
\mathrm{SSIM}(x, y) = \frac{(2\mu_x\mu_y+c_1)(2\sigma_{xy}+c_2)}{(\mu_x^2+\mu_y^2+c_1)(\sigma_x^2+\sigma_y^2+c_2)},
\label{eq:SSIM}
\end{equation}
where $x$ and $y$ are the input images, $\mu_x$ and $\mu_y$ are the means of $x$ and $y$, $\sigma_x$ and $\sigma_y$ are the standard deviations of $x$ and $y$, $\sigma_{xy}$ is the cross-covariance between $x$ and $y$, and $c_1$ and $c_2$ are constants to stabilize the division when the denominator is close to zero.

%% file: experimental_setup.tex
\section{Experimental Setup}
\label{sec:details}

\subsection{System Implementation}
\begin{figure}[!h]
    \centering
    \includegraphics[width=0.45\linewidth, height=0.45\linewidth]{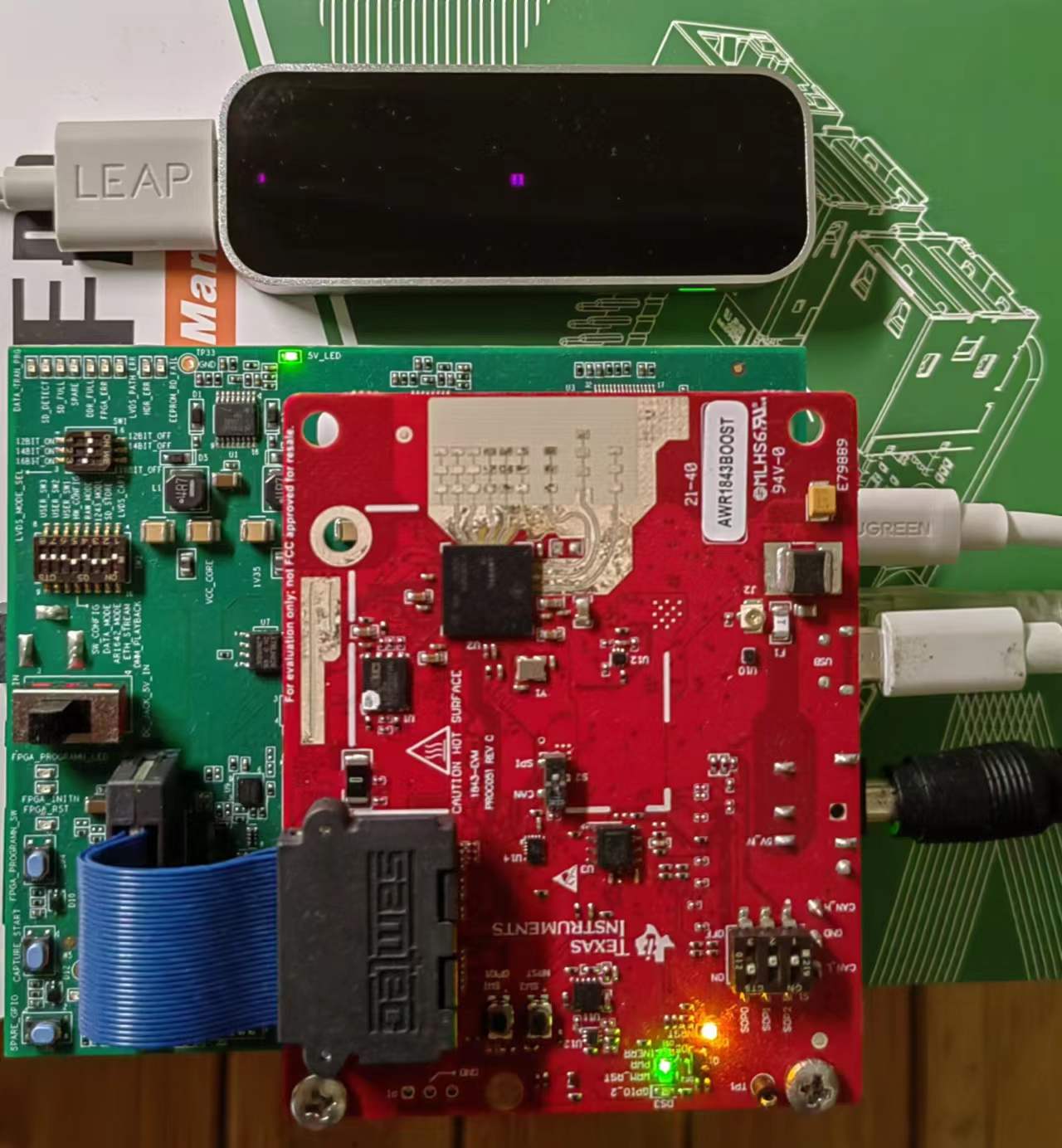}
    \includegraphics[width=0.45\linewidth, height=0.45\linewidth]{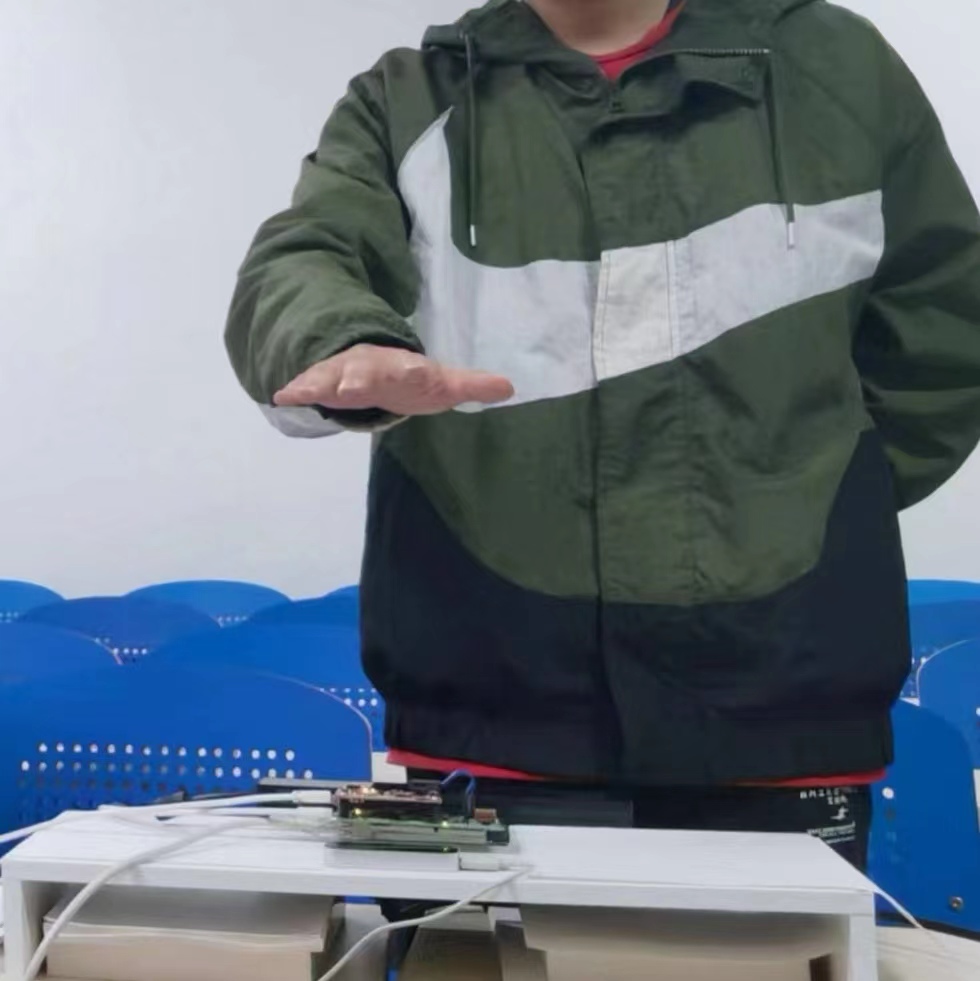}
    
    \caption{Hardware setup of data aquirsation}
    \label{fig:hardware_setup}
\end{figure}

In our gesture dataset, the radar data is collected through an AWR1843BOOST millimeter-wave radar designed by Texas Instruments (TI) Inc. We also use a TI DCA1000EVM board for real-time data capture and streaming from the radar. Only one transmitting antenna and one receiving antenna are used to send chirps with frequencies from 77GHz to 80.9GHz. The radar was set to send 10 frames per second and each frame contains 128 chirps, which consists of 256 sampling points, more detailed parameters could be seen from Table \ref{tab:radar_para}. The Leap Motion controller\cite{bachmann2018review} is a contact-free, marker-less sensor developed by Leap Motion Inc, which uses optical sensors and infrared light to detect the user's hand and finger positions in three dimensions. Fig.\ref{fig:hardware_setup}  shows the hardware setup, where the radar and Leap Motion controller are placed face up and next to each other.

\begin{table}[htb]
    \centering
    \begin{tabular}{|c|c|}
    \hline
    Parameter & Value \\
    \hline
    Starting Frequency   &  \SI{77}{\giga\hertz}              \\
    \textbf{Frequency Slope}   &      \SI{76.22}{\mega\hertz\per\micro\second}          \\
    \textbf{Bandwidth}   &            \SI{3.9}{\giga\hertz}    \\
    \textbf{Sampling Rate}   &          \SI{12500}{\mega\second\per\second}        \\
    Frame Time   &            \SI{100}{\micro\second}     \\
    Maximum Velocity   &      \SI{2.50}{\meter\per\second}           \\
    Velocity Resolution   &   \SI{0.04}{\meter\per\second}              \\
    Maximum Range   &          \SI{19.66}{\meter}           \\
    Range Resolution   &      \SI{0.38}{\meter}          \\
    Rx Gain    & \SI{30}{dB}          \\
    Tx Gain    & \SI{8}{dB}          \\
    Transmission Power  &\SI{12}{dBi}          \\
    \hline
    \end{tabular}
    \caption{FMCW Radar Parameters}
    \label{tab:radar_para}
\end{table}

\begin{figure}[!h]
    \centering
    \includegraphics[width=3.5in]{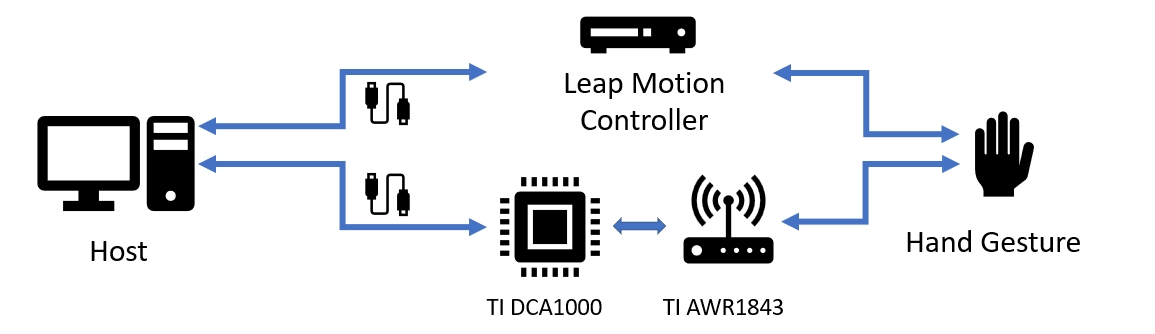}
    \caption{Data collection system architecture}
    \label{fig:collection_system}
\end{figure}

\subsection{Dataset}


\begin{table*}[ht]
\centering
\begin{tabular}{|c|l|c|c|c|c|}
\hline
\textbf{Index} & \textbf{Gesture}       & \textbf{Number of Hands} & \textbf{Occlusion} & \textbf{Angle} & \textbf{Description} \\ \hline
1 & Grasp                  & Single                   & \textbf{With}               & 0, +30, -30    & Transition from an open palm to a closed fist. \\ \hline
2 & Finger Friction        & Single                   & \textbf{With}               & 0, +30, -30    & Sliding the thumb across the index finger. \\ \hline
3 & Finger Wave            & Single                   & Without            & 0, +30, -30    & Simulating piano playing movements with all fingers over the sensor. \\ \hline
4 & Circle                 & Single                   & \textbf{With}               & 0, +30, -30    & Tracing a circular motion with the fingers above the sensor. \\ \hline
5 & Slide                  & Single                   & Without            & 0, +30, -30    & Moving the hand horizontally from right to left across the sensor. \\ \hline
6 & Finger Cross           & \textbf{Double}                   & \textbf{With}               & 0, +30, -30    & Intertwining or crossing the fingers. \\ \hline
7 & Double Hand Clap       & \textbf{Double}                   & \textbf{With}               & 0, +30, -30    & Clapping both hands together. \\ \hline
8 & Double Hand Drumming   & \textbf{Double}                   & \textbf{Without}               & 0, +30, -30    & Imitating drumming actions with both hands. \\ \hline
9 & Hand Merge             & \textbf{Double}                   & Without            & 0, +30, -30    & Joining both hands together centrally. \\ \hline
10 & Double Hand Circle     & \textbf{Double}                   & Without            & 0, +30, -30    & Each hand moving in a circular motion simultaneously. \\ \hline
\end{tabular}
\caption{Comprehensive dataset encompassing various gesture attributes, detailing the number of hands involved, the presence or absence of occlusion, the angles of capture with respect to the radar sensor, and a description of each gesture's motion and engagement of hands and fingers.}
\label{tab:gesture_attributes}
\end{table*}

In an effort to advance gesture recognition technology, we have meticulously compiled a dataset from an experiment involving 20 volunteers who were instructed to perform a sequence of 10 hand gestures commonly encountered in daily life. The dataset captures the intricate nuances of these gestures by recording each one from three distinct angles with respect to a radar sensor: directly facing it (0 degrees), and from both 30 degrees to the right (+30 degrees) and to the left (-30 degrees) of the sensor. These recordings include a wide array of movements, engaging various parts of the hands and fingers, thereby providing a rich resource for thorough analysis in gesture recognition research.

The details can be seen in Table \ref{tab:gesture_attributes}.
Structured to serve as a robust benchmark, this dataset breaks down into sub-benchmarks encompassing single-handed gestures with and without occlusion, as well as complex dual-hand interactions, each offering unique challenges to recognition systems. By doing so, it not only tests the limits of current technologies in terms of accuracy and flexibility under varied conditions but also highlights the dataset's uniqueness through its focus on practical, everyday gestures. Such a focused approach ensures that our dataset is an invaluable asset for developing recognition systems that are both effective in controlled environments and resilient in the face of real-world variability.

\subsection{Optimization Setup}
The RadarWeightNet, designed for hand gesture recognition from radar signals, was crafted using the PyTorch framework and trained on a GTX-3090 graphics card. The initial training configuration employed a learning rate of 0.0005 and a batch size of 32 for 20 epochs, with the Adam optimizer\cite{kingma2014adam}. This stage was efficiently completed in under 10 minutes. For enhanced performance, subsequent training was conducted with a lower learning rate of 0.0001, a reduced batch size of 16, and extended over 50 epochs to ensure more thorough learning.

%% file: results.tex
\section{Results}


This section presents a comprehensive evaluation of our proposed hybrid spectrum synthetic framework for radar-based hand gesture recognition. Our assessment is structured into two main parts: \textbf{evaluation of spectrum synthesis} and \textbf{evaluation of hand gesture recognition performance.}

In Section \ref{sec:v.a.}, we conduct both quantitative and qualitative analyses of our spectrum synthesis method. The quantitative evaluation compares our RadarWeightNet-enhanced approach with a baseline pure modeling method, using metrics such as Structural Similarity Index Measure (SSIM) and pixel-wise loss. We also assess the framework's adaptability across various out-of-distribution scenarios, including gesture diversity, angle variations, occlusion levels, and inter-subject variability. Qualitatively, we visually examine the synthesized spectrums, focusing on improvements in frequency shift accuracy, spectral clarity, and temporal consistency.

In Section \ref{sec:v.b.}, our evaluation focuses on the framework's effectiveness in enhancing hand gesture recognition, particularly in scenarios with limited real radar data and few-shot learning paradigms. We investigate both in-distribution and out-of-distribution settings. In the in-distribution context, we compare classification models trained on real data alone, real data augmented with modeling-based synthetic data, and real data supplemented with our RadarWeightNet-enhanced synthetic data. For the out-of-distribution scenario, we explore the potential of transfer learning by pre-training on external vision datasets, even when gesture classes don't overlap between modalities.

\subsection{Evaluation on Spectrum Synthesis}
\label{sec:v.a.}
\subsubsection{\textbf{Quantitative Evaluation}}
Our quantitative evaluation of the spectrum synthesis method focuses on both the overall performance and generalization ability. We first compare our RadarWeightNet-enhanced approach against the baseline model with pure modeling methods, demonstrating improvements in SSIM and pixel-wise loss metrics. Then we further demonstrate our hybrid framework adaptability across various out-of-distribution scenarios, including gesture diversity, angle variations, occlusion levels, and inter-subject variability, highlighting its potential for real-world applications.

To enable a quantitative assessment of our method, we leverage the Structural Similarity Index Measure (SSIM)\cite{brunet2011mathematical} expressed as in Equation \ref{eq:SSIM} and pixel-wise loss which is specifically defined as the mean squared error (MSE)\cite{wang2009mean} between the synthesized and the original spectrums, expressed mathematically as:
$L_{\text{pixel}}(m, \hat{m}) = \frac{|m - \hat{m}|^2}{N}$
where $m$ represents the original measured spectrums, $\hat{m}$ is the enhanced output, and N is the total number of pixels in a spectrum.

\begin{table}[htb]
    \centering
    \begin{tabular}{|c|c|c|c|}
    \hline
    Methods  & SSIM & MSE \\
    \hline
    Modeling   & 61.3   &   0.0076       \\
    \hline
    Modeling+RadarWeightNet   & 63.0  &   0.0072   \\
    \hline
    Improvements & \textbf{1.7} & \textbf{0.0004} \\
    \hline
    \end{tabular}
    \caption{Synthetic Spectrogram Quality Metrics: SSIM and pixel-wise loss between synthesized and original radar spectrograms. The results demonstrate the effectiveness of the Modeling-based approach and the subsequent improvements achieved by integrating RadarWeightNet, which refines the simulations and reduces the gap between synthetic and real radar data.}
    \label{tab:Image similarity}
\end{table}

\noindent \textbf{Overall performance}
As shown in Table \ref{tab:Image similarity}, the baseline method, solely reliant on modeling, achieved an SSIM of 61.3 and a pixel loss of 0.0076. However, when we augmented this method with RadarWeightNet, our enhanced neural network model, there was a marked improvement in the quality of the generated spectrums. Specifically, the integration of RadarWeightNet resulted in a notable increase in SSIM to 63.0, indicating a closer structural resemblance to the original radar data. Concurrently, the pixel loss was reduced to 0.0072, suggesting a more accurate pixel-level representation. It is worth noting that the table presents a comparison between only two methods because, to the best of our knowledge, the proposed framework is the first radar spectrum synthetic approach specifically designed for hand gesture recognition. Our synthetic methods are built upon a strong physical interpretation using cylinder-based modeling and simulation, with a neural network (RadarWeightNet) fine-tuning the weights of different reflection points. 

Regarding the comparison with generative model-based methods, we conducted comprehensive experiments using three prominent approaches: conditional generative adversarial network (CGAN), conditional variational autoencoder (VAE), and Conditional Denoising Diffusion Probabilistic Models (CDDPM). Despite our efforts, these methods failed to converge on our dataset, producing either meaningless noise or outputs lacking discernible class differences. It's important to note that direct comparison with these methods in our context is challenging due to the absence of corresponding real radar data. Consequently, the evaluation would solely rely on classification performance, which is not feasible given the non-convergence of these models. This outcome underscores the unique challenges posed by radar data synthesis and highlights the necessity for specialized approaches like our proposed framework.

\noindent \textbf{Generalization Ability}
To further evaluate our framework's performance and generalization capabilities, we implemented a comprehensive sub-evaluation strategy focusing on the adaptability across various out-of-distribution (OOD) scenarios, which we categorized into four distinct areas: variation in gesture angles, involvement of single or double hands, presence of different occlusion levels, different subjects.
Specifically, when generalizing from single-hand to double-hand gestures, our approach showed a 0.7 improvement in SSIM scores. For generalizing to occluded gestures, we observed a 1.0 enhancement in performance. Similarly, when tested on gestures at a 30-degree angle compared to the 0-degree training data, our approach exhibited a 1.0 improvement in accuracy. Even when tested on different individuals, our method demonstrated better generalization with a 0.5 improvement.
These results underscore the robustness of our RadarWeightNet-enhanced synthetic data approach in handling diverse unseen hand gestures. The consistent improvements across various scenarios affirm the effectiveness. This comprehensive evaluation strategy, encompassing gesture angle variations, hand numbers, occlusion levels, and different subjects, provides a rigorous validation of our model's generalization capabilities, particularly in scenarios involving unseen gestures.

\begin{figure}
    \centering
    \includegraphics[width=1\linewidth]{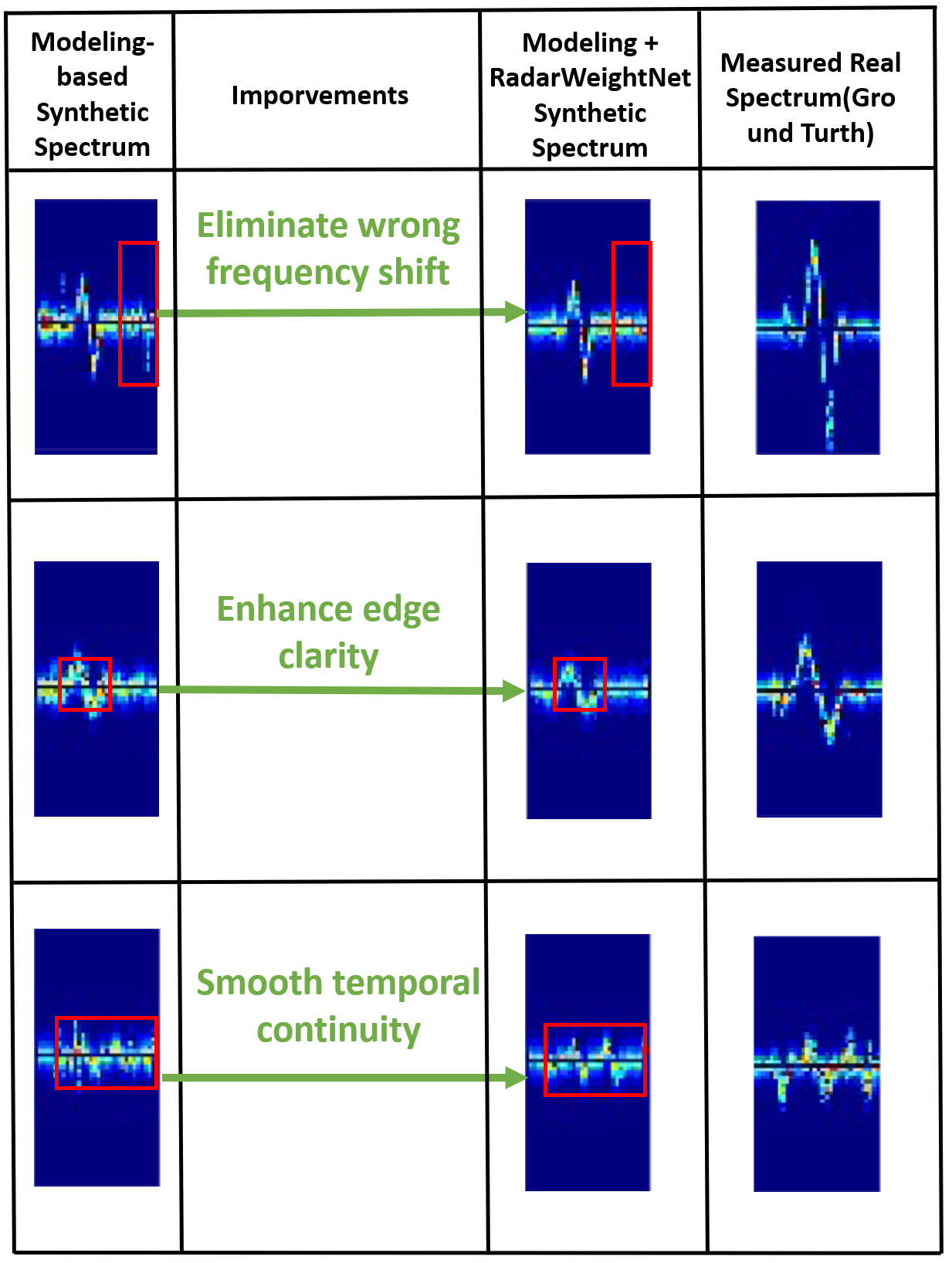}
    \caption{Visualization of the spectrum and three types of improvements.  }
    \label{fig:spectrum comparison}
\end{figure}

\subsubsection{\textbf{Qualitative Evaluation}}
Additionally, the qualitative enhancements are visually discernible in the spectrums themselves.
Our RadarWeightNet demonstrates significant improvements in three key areas: elimination of erroneous frequency shifts, enhancement of spectral clarity, and improvement of temporal consistency. As illustrated in Figure. \ref{fig:spectrum comparison}, these improvements address common limitations of existing modeling-based methods.
Firstly, RadarWeightNet enhances the accuracy and clarity of the synthetic spectrum, particularly in scenarios where visual input may be unreliable. Traditional modeling-based approaches often generate inappropriate frequency shifts or blurred spectral edges due to inaccurate skeleton estimation or occlusions in the visual data. Our approach mitigates these issues by identifying problematic reflection points and adaptively weighting their contributions to the final spectrum. This mechanism not only corrects erroneous frequency shifts but also sharpens spectral representations, thereby improving the overall fidelity of the synthetic radar data.
Also, RadarWeightNet improves the temporal consistency of the synthetic spectrum. Our method produces more coherent and realistic temporal sequences, which is crucial for accurately representing the continuous nature of hand gestures in radar data.
These enhanced patterns are anticipated to be advantageous in subsequent applications, facilitating improved gesture recognition performance.



\subsection{Evaluation on Hand Gesture Recognition}
\label{sec:v.b.}
This subsection evaluates the effectiveness of our hybrid spectrum synthetic framework in enhancing hand gesture recognition through the utilization of synthesized spectrums from vision modalities, with particular emphasis on scenarios characterized by limited real radar data and few-shot learning paradigms. In subsection \ref{sec:v.b.1}, we initially assess the framework's performance within an in-distribution context, where gesture classes are congruent across video and radar datasets. Our experimental methodology involves a comparative analysis of classification models trained on three distinct dataset configurations: exclusively real data, real data augmented with modeling-based synthetic data, and real data supplemented with our novel RadarWeightNet-enhanced synthetic data. Subsequently, subsection \ref{sec:v.b.2} extends our investigation to out-of-distribution scenarios, examining the potential for transfer learning by pre-training on external vision datasets. This approach aims to determine whether such pre-training can confer benefits to radar gesture recognition, even in the absence of overlapping gesture classes, thereby demonstrating the framework's versatility and its capacity for generalization in real-world applications.

\subsubsection{Classification under in-distribution Setting}
\label{sec:v.b.1}

In the in-distribution setting, where video and radar data share identical gesture classes, we leverage the abundance of video data to generate corresponding radar spectra for common gestures. This approach expands the radar dataset scale, thereby enhancing recognition performance, particularly in radar-only systems deployed for privacy protection. To be specific, we leveraged our multi-modal HGR dataset and employed the ResNet50 architecture \cite{he2016deep} as the foundational classification model and then the model's performance across three dataset variations: (i) an original subset consisting of n\% real radar data, (ii) the aforementioned real radar data supplemented with modeling-based synthetic data, and (iii) the aforementioned real radar data supplemented with RadarWeightNet-Enhanced Synthetic Data. 
To simulate environments with severe radar data scarcity, we systematically reduced the volume of real radar data from 50\% to 5\%. A consistent partitioning strategy was adopted for testing, maintaining a 20\%-80\% split for the genuine data subset and a corresponding split for the combined datasets.  \textbf{To have a comprehensive evaluation, we also tried generative-based methods as a substitution for the RadarWeightNet-Enhanced Synthetic Data using the Conditional Generative Network as the backbone \cite{mirza2014conditional}, but the model was hard to converge with such little data to train since our scenario is a few-shot manner.}

The classification outcomes, as exhibited in Table \ref{tab:classification_result}, reveal a pronounced elevation in accuracy when the training set is expanded with synthetic data, particularly in cases where real data is sparse. When only 5\% of the real radar data is used, the accuracy improves from 44.4\% to 88.9\% (an improvement of 44.5 percentage points) with the addition of RadarWeightNet-Enhanced Synthetic Data. Similarly, with 20\% and 50\% of real radar data, the accuracy increases by 30.6 and 19.5 percentage points, respectively. Our tailored RadarWeightNet model demonstrated exceptional proficiency in emulating genuine radar signal characteristics, which in turn translated to superior classification performance compared to the generative-based methods. 

\begin{figure}[htbp]
    \centering
    \includegraphics[width=0.9\linewidth]{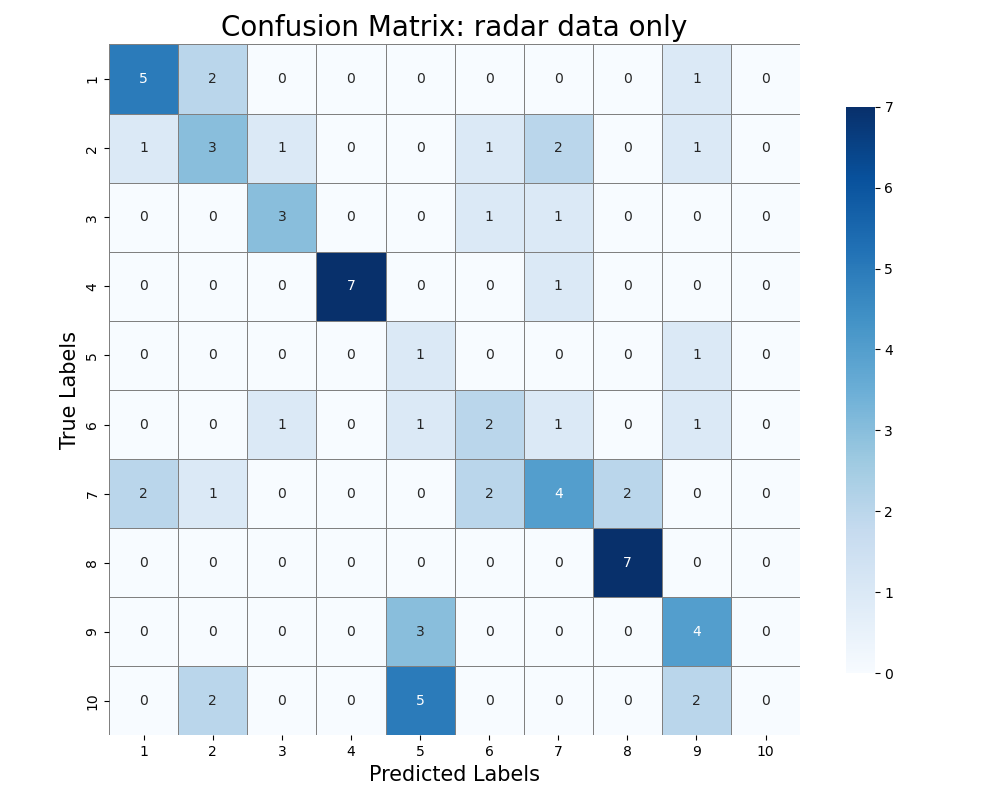}
    \includegraphics[width=0.9\linewidth]{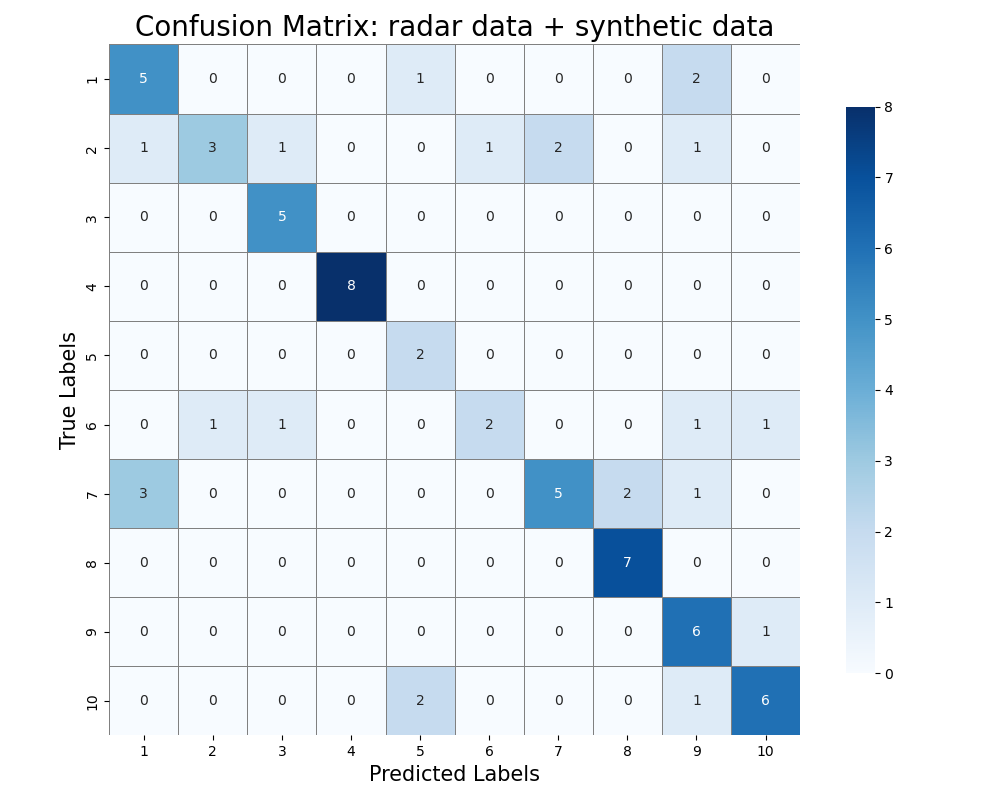}
    \caption{Confusion matrix on gestures from vision dataset, the one above is trained with pure measured data and the other is trained with measured data and enhanced synthetic data(The percentage is 20\%).}
    \label{fig:confusion_matrix}
\end{figure}

The advantages are further highlighted through the confusion matrix depicted in Figure \ref{fig:confusion_matrix}. All the gestures share the same index as Table \ref{tab:gesture_attributes}. Specifically, all the class accuracies have been improved, especially for gestures 10, corresponding to the \textit{Double Hand Circle} gesture. This poses a significant challenge for radar-based recognition, which is the gestures primarily generate tangential velocity resulting in less informative Doppler spectra compared to gestures with more pronounced radial movements, as radar systems mainly detect radial velocity. Our synthetic data augmentation framework demonstrates significant improvement in recognizing this challenging gesture.

\begin{table*}[htb]
    \centering
    \renewcommand{\arraystretch}{1.2}  
    \begin{tabular}{|l||ccc|c|c|}
    \hline
    \textbf{Training Data} & \textbf{Radar} & \textbf{Radar +} & \textbf{Radar +} & \textbf{Improvement} & \textbf{Generative} \\
    \textbf{Percentage} & \textbf{Only} & \textbf{Model-based} & \textbf{Enhanced} & & \textbf{Model} \\
    & & \textbf{Synthetic} & \textbf{Synthetic} & & \\
    \hline
    5\%   & 44.4  & 83.3 & \textbf{88.9} & 44.5 & N.A. \\
    20\%  & 55.5  & 83.3 & \textbf{86.1} & 30.6 & N.A. \\
    50\%  & 68.3  & 82.2 & \textbf{87.8} & 19.5 & N.A. \\
    \hline
    5\%*  & 33.6  & 80.8 & \textbf{81.1} & 47.5 & N.A. \\
    20\%* & 64.2  & 80.5 & \textbf{81.9} & 17.7 & N.A. \\
    50\%* & 73.3  & 81.4 & \textbf{82.0} & 8.7  & N.A. \\
    \hline
    \end{tabular}
    \caption{Performance comparison of HGR models with different training strategies. Results show accuracy (\%) for models trained with varying amounts of real radar data, with and without synthetic data augmentation. The addition of enhanced synthetic data significantly improves performance, especially in low-data scenarios (5\% real data). Asterisk (*) indicates results on test sets split from the complete dataset. Generative models (CGAN, DDPM, VAE) failed to converge on our dataset.}
    \label{tab:classification_result}
\end{table*}

\subsubsection{Classification under OOD Setting}
\label{sec:v.b.2}


Previous evaluations have primarily focused on in-distribution scenarios, where hand gestures are common and have corresponding representations in the vision dataset. However, real-world applications often encounter out-of-distribution (OOD) situations, where certain hand gestures are not included in the training data. This raises an important question:
\textbf{Can our radar HGR model benefit from pre-training on any external vision dataset, even when there's no overlap between the pre-trained spectrum dataset and the radar gesture classes?}
To address this challenge, we explore the potential of transfer learning in an OOD setting. Specifically, we investigate whether a model pre-trained on vision data can effectively adapt to recognize previously unseen radar gestures. We employ a few-shot learning approach, where we initially train our classification model using the comprehensive vision dataset. Subsequently, we fine-tune this pre-trained model using a very small radar dataset containing the new, unseen gestures. This methodology allows us to assess the model's ability to generalize and quickly adapt to novel gesture classes, thereby evaluating its robustness in real-world scenarios where new, unanticipated gestures may emerge.

\begin{table}[htb]
    \centering
\begin{tabular}{|c|c||c|c|}
    \hline
    Gesture Classes & Real Data Percentage & w/o pre-train & pre-trained \\
    \hline
    \multirow{2}{*}{10} & 5\% & 43.3\% & \textbf{51.1\%} \\
    \cline{2-4}
    & 10\% & 54.7\% & \textbf{55.0\%} \\
    \hline
    \multirow{2}{*}{5} & 5\% & 61.0\% & \textbf{67.8\%} \\
    \cline{2-4}
    & 10\% & 72.7\% & \textbf{76.7\%} \\
    \hline
\end{tabular}
    \caption{This table compares classification accuracies using a ResNet50 model pre-trained on the unpaired vision-generated spectrum and fine-tuned with limited radar data. The results, categorized by gesture class number and data percentage, demonstrate the benefits of cross-modality pre-training in few-shot learning scenarios, even though there is no overlapping class.}
    \label{tab:pre-train}
\end{table}

\noindent \textbf{External Vision Dataset}
To answer this, we tested on the DHG 14/28 \cite{smedt2016dynamic}, a vision-based hand gesture dataset that utilizes a depth camera to capture detailed information on 22 hand joints during various hand gestures. The dataset comprises 14 distinct hand gestures, each performed five times by 20 participants. The gestures included in the dataset are: Grab, Tap, Expand, Pinch, Rotation CW, Rotation CCW, Swipe Right, Swipe Left, Swipe Up, Swipe Down, Swipe X, Swipe V, Swipe +, and Shake. With the exception of the Grab gesture, all other gestures in this dataset differ from those in our study, as outlined in Table \ref{tab:gesture_attributes}.

\noindent \textbf{Skeleton Alignment and Spectrum Generation:}
Apart from the hand gesture class, the 3D hand skeleton model is also different from ours, where the 22 joints of the hand skeleton returned by the Intel RealSense camera include 1 for the center of the palm, 1 for the position of the wrist, and 4 joints for each finger representing the tip, the 2 articulations, and the base. 
We make slight adjustments to ignore the center of the palm and merge the two articulations of the thumb. Also, we transform the 3D coordinates system to align the difference in the y-axis and z-axis, then translate all the gestures to the approximate place where there is a similar distance from the sensor to our aiming scenario. 
Then, based on the 3D coordinate sequences, we calculate the physical characters of the motion for each gesture and feed these into our synthetic framework to generate the corresponding spectrum. 

\noindent \textbf{How pre-trained and few-shot learning helps:}
To evaluate how our synthetic model could leverage the external vision dataset to help the radar hand gesture recognition, we train the model in a pre-train and few-shot learning manner using the same network architecture. Firstly, we change the fully connected layer's output into 14 to correspond to the 14 class of gestures in DHG 14/28 dataset. After that, we train the network to classify different synthetic spectrums during which we save the best model for the following. 
Then, we change the fully connected layer's output into 10 then fine-tune the model with only 5\% or 10\% percentage of our measured radar data. Table \ref{tab:pre-train} demonstrates the effectiveness of our pre-training approach using vision-generated synthetic data for radar-based gesture recognition. Pre-trained models consistently outperform those without pre-training, with the fully fine-tuned pre-trained model achieving the best results across all scenarios. The improvements are particularly significant in few-shot learning settings, especially when there are more classes.

\section{Analysis}
In this section, we first demonstrate the generlization ability on four key factors that impact the performance when transferring the pre-trained model to real radar data and based on that how to augment the vision data to ease the OOD accordingly. After that, we further demonstrate how and why the RadarWeightNet works especially for the HGR setting, and end with some ablation studies on the RadarWeightNet.



\subsection{Why RadarWeightNet: Optimizing Reflection Point Contributions}

Accurately simulating hand gesture radar signals is challenging due to the complex interplay of multiple atomic structural patterns and distinct movements of various hand parts. Our approach with RadarWeightNet offers a unique balance between data-driven adaptation and physical interpretability by decomposing a hand gesture's radar spectrum into constituent components from individual reflection points. This decomposition reveals how different parts of the hand contribute to the overall synthetic spectrum, providing crucial insights into the sources of discrepancies between synthetic and real radar spectrums.

\begin{figure}[htbp]
	\centering  
		\includegraphics[width=0.16\linewidth]{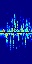}
		\includegraphics[width=0.16\linewidth]{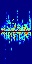}
		\includegraphics[width=0.16\linewidth]{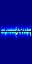}
            \includegraphics[width=0.16\linewidth]{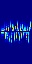}
            \includegraphics[width=0.16\linewidth]{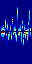}
	\caption{Spectrum Decomposition by Reflection Points: The first two images show the real and synthetic radar spectra, with subsequent images detailing the synthetic spectrum decomposed into individual reflection points. This nuanced breakdown aids RadarWeightNet in pinpointing and correcting variances between synthetic and real data, thereby improving gesture recognition fidelity. }
 \label{fig:Decomposition_points}
\end{figure}

As illustrated in Figure \ref{fig:Decomposition_points}, by visualizing the frequency shifts corresponding to each reflection point, we observe varying levels of accuracy in their contributions. Some points generate signals that closely match the real spectrum, while others introduce noticeable errors. For instance, in a finger wave gesture, the spectrum components from the static palm might align well with real data, while those from moving fingertips might show discrepancies due to modeling inaccuracies or occlusion effects.
RadarWeightNet addresses these discrepancies by intelligently adjusting the weights assigned to each reflection point's signal, amplifying accurate contributions while attenuating erroneous ones. This process compensates for modeling errors, accounts for occlusion effects, and captures complex inter-reflector interactions. By refining the synthetic spectrum while maintaining interpretability, RadarWeightNet enhances the overall fidelity of the synthetic spectrum, achieving a balance between data-driven learning and physics-based decomposition.


\subsection{Generalization Capabilities of the HGR Model}
\label{sec:vi.a.}
Similar to our earlier spectrum synthesis assessment, we build our evaluation of the enhanced hand gesture recognition model's generalization capabilities in out-of-distribution (OOD) scenarios, focusing on gesture angle variations, different subjects, single vs. double-handed gestures, and varying occlusion levels.
We leveraged our RadarWeightNet to generate synthetic radar spectrum data from visual sources, addressing the challenge of radar data scarcity. As shown in Table \ref{tab:zero-shot}, training with 20\% real radar data supplemented by 80\% synthetic data significantly outperformed training with 20\% radar data alone across all OOD scenarios. Notable improvements were observed in generalization across different subjects (28.0\% increase) and gesture angle variations (21.1\% increase). Smaller but significant gains were also seen in hand number variations (3.9\% improvement) and occlusion levels (3.3\% increase). These results demonstrate our RadarWeightNet-enhanced approach's robustness in handling diverse unseen hand gestures, effectively expanding the scope of few-shot learning and ensuring a more generalizable model despite limited radar data.

\begin{table}[htb]
    \centering
    \begin{tabular}{|c|c|c|c|c|}
    \hline
        Scenarios & Angles & Subjects & Hand Numbers & Occlusion  \\
    \hline
       \parbox{1.5cm}{\vspace{0.15cm} train with 20\% radar data \vspace{0.15cm}} & 58.0\% & 51.1\% & 71.6\% & 71.1\%  \\
    \hline
        \parbox{1.5cm}{\vspace{0.15cm}train with 20\% radar data + 80\% synthetic data\vspace{0.15cm}}  & \textbf{77.7\%} & \textbf{79.1\%} & \textbf{75.5\%} & \textbf{74.4\%} \\
    \hline    
        Improvements & 21.1\% & 3.9\% & 3.3\% & 26.6\% \\
    \hline
    \end{tabular}
    \caption{Assessment of Classification Model Generalization on OOD Scenarios: This table compares the baseline generalization capacity with the RadarWeightNet-enhanced performance across various OOD scenarios. The results demonstrate the efficacy of RadarWeightNet in improving the robustness of gesture recognition models to OOD instances, with marked performance gains in all tested categories. This underscores the benefit of incorporating synthetic radar data to bridge the gap in radar data availability and enhance few-shot learning. }
    \label{tab:zero-shot}
\end{table}

\subsection{Assessment of Physical Characteristics and Network Complexity in RadarWeightNet}
In our comprehensive evaluation, we have quantitatively assessed the enhancement in the Structural Similarity Index Measure (SSIM) brought about by the integration of diverse physical features into RadarWeightNet. These features include visibility, Radar Cross Section (RCS), distance, velocity, and acceleration, extracted from coordinates obtained via the Leap Motion controller or similar vision sensors. The data, detailed in Table \ref{tab:Ablation Study}, indicates that although the improvements are subtle, the inclusion of specific physical characteristics does indeed lead to a modest yet noticeable increase in performance.

\begin{table}[htb]
    \centering
    \begin{tabular}{|c|c|c|c|c|c|}
    \hline
        Feature & visibility & RCS & distance & velocity & acceleration\\
    \hline
        Improvement &  1.63\%  & 1.66\% & 1.63\% & 1.65\% & 1.67\%\\
    \hline
    \end{tabular}
    \caption{Impact of Physical Feature Integration on SSIM in RadarWeightNet: This table presents the percentage improvements in SSIM achieved by incorporating various physical features into RadarWeightNet, underscoring the beneficial yet incremental enhancements in radar signal fidelity for gesture recognition.}
    \label{tab:Ablation Study}
\end{table}

In conjunction with this study, we have also explored the implications of adding complexity to the network's architecture. Through further experimentation, it has been shown that RadarWeightNet is capable of efficiently learning invariant features from the physical attributes without resorting to more complex network designs. The network’s ability to quickly converge and maintain high performance, even when trained with a sparse dataset, suggests that a simpler, more streamlined architecture suffices. This finding aligns with the marginal differences observed in the SSIM improvements, reinforcing the idea that sophisticated network structures are not necessarily beneficial for the task at hand. The balance between the input of physical characteristics and the network's structural simplicity forms the crux of RadarWeightNet's effectiveness in radar-based hand gesture recognition.

%% file: conclusion.tex


\section{Conclusion}
In this paper, we present a novel hybrid spectrum synthetic framework for advancing radar-based hand gesture recognition. Our approach addresses key limitations of existing methods by combining a physically grounded hand reflection model with a data-driven neural network optimization. This methodology enables data-driven signal fine-tuning while maintaining physical interpretability. A key takeaway from our research is the importance of decomposing and analyzing the impact of individual reflection points on the overall spectrum. By examining the combined relationships of signals from different reflection points, we developed a fine-tuning technique that offers a new perspective on integrating neural networks and physical simulations. Crucially, our framework successfully addresses the convergence challenges faced by generative model-based methods when radar data is extremely limited, demonstrating effective synthetic radar spectra generation even in scenarios where real radar data is scarce.

Our comprehensive evaluation, utilizing a multi-modal dataset comprising 10 distinct hand gestures, convincingly demonstrates the efficacy of our approach. The integration of RadarWeightNet yielded significant improvements in synthetic spectrum quality, elevating the SSIM to 63.0 while reducing pixel loss to 0.0072. This enhancement in spectrum generation translates into substantial gains in gesture recognition accuracy, particularly evident in few-shot learning scenarios. Notably, our method effectively leverages external vision datasets to bolster radar-based recognition, even in the absence of overlapping classes. For instance, with a mere 5\% of real radar data for 10 gesture classes, our pre-trained model achieved an impressive 18\% relative improvement in accuracy. This capability opens new avenues for data augmentation in radar-based gesture recognition. Our work marks a significant advancement in radar spectrum synthesis for hand gesture recognition, offering a robust solution to the persistent challenges of limited radar datasets and complex hand-level interactions.